\documentclass{sig-alternate-05-2015}

\PassOptionsToPackage{hyphens}{url}\usepackage[draft]{hyperref} % Required for hyperlinks
\usepackage{listings}
\usepackage{amsfonts}
\usepackage[latin1]{inputenc}
\usepackage[english]{babel}
\usepackage{amsmath}
\usepackage{booktabs}
\usepackage{boldline}
\usepackage{tabularx}
\usepackage{fancyvrb}
\usepackage{algorithm,algorithmic}
\usepackage{chngcntr}
\usepackage{subfigure}
\usepackage{url}

\begin{document}\sloppy

\title{How a simple bug in ML compiler could be exploited for backdoors?}

\numberofauthors{1}

\author{
\alignauthor
Baptiste David\\
\affaddr{\'{E}cole nationale supérieure d'arts et métiers (ENSAM)}\\
\affaddr{\'{E}cole d'ingénieurs du monde numérique (ESIEA)}\\
\affaddr{Laboratoire de cryptologie et virologie opérationnelles --- $\left(\text{C} + \text{V}\right)^{\text{O}}$}\\
\affaddr{38 rue des Docteurs Calmette et Guérin 53000 Laval, France}\\
\email{baptiste[dot]david[at]ensam[dot]fr}
}

\maketitle

\abstract{Whenever a bug occurs in a program, software developers assume that the code is flawed, not the compiler. In fact, if compilers should be correct, they are just normal software with their own bugs. Hard to find, errors in them have significant impact, since it could result to vulnerabilities, especially when they silently miscompile a critical application. Using assembly language to write such software is quite common, especially when time constraint is involved in such program.

This paper exposes a bug found in Microsoft Macro Assembler (ml for short) compiler, developed by Microsoft since 1981. This assembly has the characteristics to get high level-like constructs and high level-like records which help the developer to write assembly code. It is in the management of one of this level-like construct the bug has been found.

This study aims to show how a compiler-bug can be audited and possibly corrected. For application developers, it shows that even old and mature compilers can present bugs. For security researcher, it shows possibilities to hide some unexpected behavior in software with a clear and officially non-bogus code. It highlights opportunities for including stealth backdoors even in open-source software.}
 

\section{Introduction}

Compilers are piece of software designed to build new software according to their source code. It exists a lot and most famous ones have made most famous software. It exists compilers for every compiled language, such as C\#{}, C++, C and assembly. The last is one of the eldest one, not often used except in critical projects where efficiency and low design consideration must be strongly taken into account.\newline

Write assembly code directly is reputed to be hard and complicated. Every facility provided by more abstract languages are not present and implementing simple algorithms can be a complicated task for developers. There exists different assembly syntaxes (AT\&{}T and Intel for the most famous ones) and different languages, such as GoAsm, NASM or MASM. The last one, MASM \cite{MASM} which stands for Microsoft Macro Assembler \cite{MasmArticle}, is famous by the ease it provides, since 1981, to implement programs thanks to the use of its macro code \cite{MasmArticleBis}. If this ease helps everyday development, this one must be reliable since it rules what programs are supposed to do. In other word, the compiler must not perform any error since these ones are very complicated to find. Indeed, when a bug occurs, the first reflex is about to search from source code and not blaming the compiler. Source code defines the logic the program is supposed to follow. And error coming from a correct source code which would result in an unexpected behavior could be very hard to find. Compilers are supposed to be reliable since software logic depends on them.\newline

Compilers bugs are more common than what is usually thought \cite{WikiBugs, CompilerBug}. A paper \cite{GccLlvmStat} about statistics on bugs' compilers examined about five thousand ones on a decade's span. Technical explanations about such bugs had already been performed for C language \cite{BugCPaper, WarningBugs}, for instance. Compilers' bugs come from implementation erros, undifined behaviour in programming languages or unfortunate optimization made by the compiler \cite{OptimizeBugs}. When such errors are found, they must be reported as soon as possible since software can be targeted by unexpected logic problems. But, it could be interesting to imagine how an attacker could exploit such a default. Indeed, open source software are supposed to be secure because everyone can check the source code for unexpected behavior. However an attacker, from a full legitimate open source code, could use a flaw in the compiler to run, at execution time, unexpected features in contradiction of what was originally written in the source code. Different from Unix backdoor \cite{LinuxBackDoor}, this one is close to what is explained in \cite{BackDoorBugCompiler}. Hidden from a trap present in the compiler (deliberatly \cite{Thompson} or not), the goal is about to hide the malicious purpose of a program despite its clean source code. Such bugs --- designed to build backdoors --- can be found with dedicated processes \cite{BugMutation} or by the developer when facing an \textit{impossible} bug in his/her program.\newline

This is the aim of the article since it proposes in section~\ref{sec:definedproblem} to show and explain a bug in \textbf{ml} compiler --- masm compiler coming from Microsoft Visual Studio software. Then, in section~\ref{sec:HowToBuildABackDoor} would be interesting to explain how an attacker could exploit it. Finally, conclusion about this disclosure will be described.

\newpage

\section{Macro assembly and ml mistake with Boolean negation operator}
\label{sec:definedproblem}

\begin{figure}[!h]%
    \centering
    \subfigure[Simple C condition]{{\includegraphics[width=4.5cm]{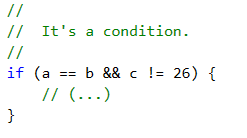} }}%
    \qquad
    \subfigure[Equivalent assembly code.]{{\includegraphics[width=\columnwidth]{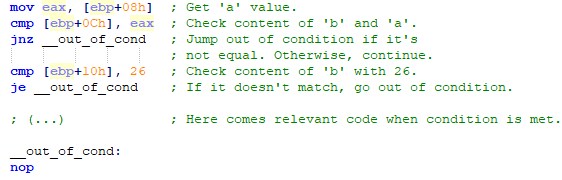} }}%
    \caption{Equivalent code in C and assembly.}%
    \label{fig:CAndAssemblySimple}%
\end{figure}

Assembly language is less complicated than people think. Not far from C language, management of the stack, calling convention and program's control flow by conditions and loops are main differences from a technical point of view. This is the last point which is interesting in our case. In C, writing conditions is quite simple and the same can be performed with assembly language, as displayed in figure~\ref{fig:CAndAssemblySimple}.

To understand the assembly code, the reader must take into account that local values are stored on the stack in x86 assembly and, in our example, are referenced from \textsc{ ebp} register used as base pointer. \textsf{Cmp} instruction can be viewed as the subtraction between two values with the content of each variable preserved (only the resulting \textsc{eflag} register is modified, according to the operation). In case of equality, the difference is zero, otherwise it is not, which explains why \textsf{jne} (jump not equal) instruction is used there.\newline

For obvious reasons, it is more interesting to write algorithms than technical codes, macro assembly such as MASM provides facilities for developers to help them in conditions writing. Something equivalent, close to C language, can be written.

\begin{figure}[!h]%
    \centering
   \includegraphics[width=\columnwidth]{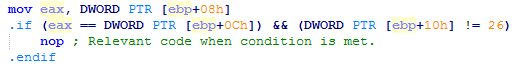}
   \caption{Macro assembly condition with MASM.}
   \label{fig:MasmCond}
\end{figure}

\begin{figure}[!h]%
    \centering
   \includegraphics[width=\columnwidth]{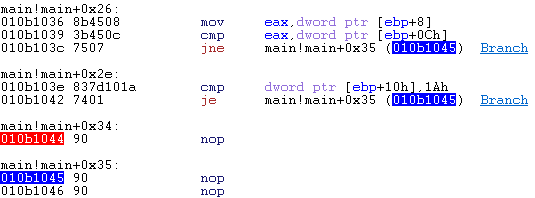}
   \caption{Disassembly of code in figure~\ref{fig:MasmCond}.}
   \label{fig:windbgmasmcond}
\end{figure}

Easier to read, useful for long development, this code is halfway between C and assembly code (note the use of brackets do not change complexity here). Compiled with \textbf{ml} and disassembled with a debugger such as \textbf{Windbg}, the result is provided in the figure~\ref{fig:windbgmasmcond}.\newline

The code is built in order to force the program to jump at \textsl{0x010b1045} if one of the condition is not met. Otherwise, the program continues its normal flow to reach \textsl{0x010b1044}. In addition to \textbf{AND} conditions, MASM can build \textbf{OR} or \textbf{NOT} operators  \cite{Masm61ProgrammerGuide}. In the following example, we build two codes to illustrate the action of the \textbf{NOT} operators from macro assembly to generated op-codes by the \textbf{ml} compiler. On the left column, we test whether the value stored in \textsc{eax} is different from zero or not. If it is not, instruction \textsf{add ebp, 4} is executed. Otherwise, we go to the following nop instruction. On the right column, the condition is negated, it means if the value stored in \textsc{eax} is zero, then, the instruction \textsf{add ebp, 4} is now executed.

\begin{figure}[!h]%
    \centering
   \includegraphics[width=\columnwidth]{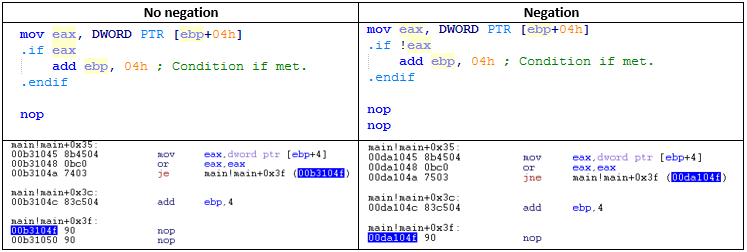}
   \caption{Effect of \textbf{NOT} operator with MASM.}
   \label{fig:NegateArray}
\end{figure}

The test operation is performed by \textsf{or eax, eax}1 instruction in both cases. This specific operation could look like a \textit{no operation} since it is not supposed to perform any modification on \textsc{eax} content. That is the goal in a sense it preserves the content of the value but set specific flags in \textsc{eflag} register to allow conditional instruction to works according to the content of \textsc{eax}, especially if \textsc{eax} is zero or not.\newline

The difference comes in the conditional jump instruction which is performed right after. In the first case, \textsf{je} which stands for "\textit{jump near if equal}" means that jump operation is performed when flag zero is set to one (ZF=1). In other word, if the result of \textsf{or eax, eax} would set ZF to one if \textsc{eax} was zero. In such a case, the jump will be executed and the instruction provided if condition is met would be skipped (going to nop instruction). Or course, if content of \textsc{eax} is different from zero, the instructions inside the if statement are executed.\newline

For the second case, the conditional jump is now \textsf{jne} which is the opposite version of \textsf{je}. The condition is met when the zero flag is set to zero (ZF=0). From a logic point of view, it is a negation of the first case. Until that point, everything works correctly as it is supposed to be.\newline

Issues come when we raise the level of complexity in conditions. Consider the following conditions:\newline

\begin{figure}[!h]%
    \centering
   \includegraphics[scale=0.85]{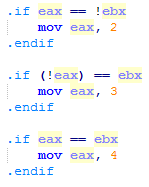}
   \caption{Use of \textbf{NOT} operator in conditions.}
   \label{fig:ComplexCondNot}
\end{figure}

For obvious reasons, all of them are supposed to produce different results, according to the values stored in \textsc{eax} and \textsc{ebx}. Technically speaking, a condition such as \textsf{(eax == !ebx)} could be rewritten as \textsf{(eax == (ebx == 0))} to be more understandable. The same applies for \textsf{((!eax) == ebx)} and the use or the lack of brackets do not change anything for the compiler. Because all conditions are different, they all should be computed differently (use of different op-codes or conditional jumps, for instance) because each describes a different condition to be met. The main question stands in the following disassembly of these conditions where all conditions are compiled in the exactly same way.

\begin{figure}[!h]%
    \centering
   \includegraphics[width=\columnwidth]{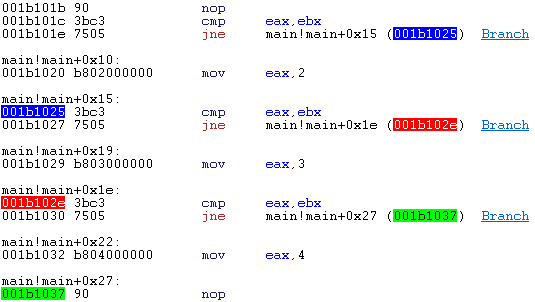}
   \caption{Disassembly of code compiled in figure~\ref{fig:ComplexCondNot}.}
   \label{fig:DebugComplexNotCond}
\end{figure}

Each condition is built in the same manner (\textsf{cmp eax, ebx} followed by a conditional jump if not equal --- \textsf{jne} --- to the next condition) despite the fact they are different. From the \textbf{ml} compiler point of view, \textsf{eax == ebx} is the same as \textsf{(!eax) == ebx}. Any debugging session is enough to confirm that the logic written originally in the source code is not respected at generation time by the compiler.\newline

To prove an error is made by \textbf{ml} compiler, we wrote an equivalent code in C, compiled it with visual studio compiler and reversed it with IDA software\footnote{\url{https://www.hex-rays.com/products/ida/}}. The C equivalent code is written with the constraint to retrieve two values from the user. Doing so, no one could object compiler could have made any optimization in the backyard.\newline

\begin{figure}[!h]%
    \centering
   \includegraphics[width=\columnwidth]{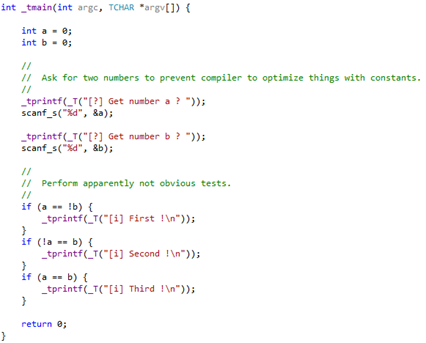}
   \caption{C code supposed to test \textbf{NOT} condition.}
   \label{fig:CTestCond}
\end{figure}

Decompiled code is a piece of cake to check how the compiler built this source code.

\begin{figure}[!h]%
    \centering
   \includegraphics[width=\columnwidth]{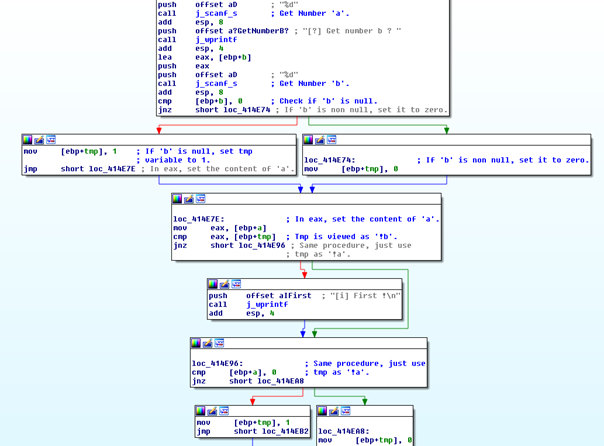}
   \caption{Decompiled code from the one compiled in figure~\ref{fig:CTestCond}.}
   \label{fig:IdaDecompiled}
\end{figure}

Difference with the code compiled with \textbf{ml} is blatant. The first block of code is dedicated to retrieve values from the user. Then, the first condition \textsf{(a == !b)} is computed by the use of a temporary value stored on the stack. Indeed, according to the value of \textbf{b}, a temporary local value is set to one or zero (second and third block) to be finally compared to a value at the end. In other words, the negation of \textbf{b} is stored in a \textit{shadow} local value to be used in the comparison next. The same applies for the next condition (in such a case, it is the value a which is used).\newline

According to what the C compiler did, there is an issue with \textbf{ml} compiler. Note C compiler uses a temporary value to compute the negation of one value, what \textbf{ml} compiler does not. An explanation could sit in the lack of capacity of \textbf{ml} to build its own local values as one developer could craft his/her own ones.\newline

Finally, \textbf{ml}'s bug resides in the use of negation operator in an equality (or inequality) test. Every code which follows the pattern \textsf{!register sign register/memory} is eligible to be incorrectly compiled. This bug is introduced by a bad management of the value negated. Indeed, the use of \textbf{NOT} operator changes the content of the value \textit{inside} the condition. Out of the condition, the negated value is not supposed to be altered. The lack of use of temporary values leads to misinterpret the condition, resulting to the bug. Note that regular negation from a single value (for instance: "\textsf{.if !eax}") or a full condition (such as: "\textsf{.if !(condition)}") work perfectly fine since it is the test instruction or the conditional jump which is impacted in that case.

\section{How to build a sneaky backdoor with ml compiler bug?}
\label{sec:HowToBuildABackDoor}

\subsection{Context of the backdoor}

If it is possible to exploit the bug to execute unexpected code despite official logic of the source code, a backdoor is writable. The goal is to get an official source code which should pass a code review from anyone (expert or not). Of course, reverse engineering compiled code from application would be enough to find the \textit{mistake} we included in the logic flow of the process... But who cares about reversing open-source application that you can compile by yourself ? Of course, because we modify original source code, source code of the targeted application must be accessible. Open-source software are perfect target for that.\newline

As described in \cite{BackDoorBugCompiler}, two plots are available to build the backdoor. First, you own the project you want to backdoor. On the positive side, it makes it easy to introduce the backdoor since you are the main leader developer. On the negative side, if the backdoor is found, you lose and here comes consequences on your reputation. To reduce the possible cost, the second plot is about to propose a patch for an existing software. It allows you to patch \textit{several} open-source projects since you can act as an occasional contributor. It is stealthier, in the sense you are not on first line. In addition, you get the benefit to possibly argue it was a plausible mistake instead of a malicious intent by design. In all cases, attack is performed through a modification of an existing software to detour it is original procedure.\newline

The main part, about the modification we submit, is about to not raise a security issue. First, no inspection (from human to full formal verification) of code would be able to detect a security breach in the code since the official security logic is respected. Then, the bug we exploit avoids questions about a dubious (or, at best, a nonsense) functionality inserted in code. After all, we are fooling the logic of conditions in a sneaky way (a condition supposed to be false is going to be interpreted finally, as true) so that code, which is supposed to reject something, now, accepts it.\newline

Even if the code is detected, it is still possible to deny any responsibility in the vulnerability found. The logic of checks in the original code is supposed to be true and you were not supposed to know the debugger had a bug. At worst, one can blame you not to have tested your software enough. Of course, with an example as simple as the one provided in figure~\ref{fig:exampleBDI}, it can be complicated to argue we did not test it. But more complex cases can be implemented.

\subsection{Description of bug consequences for backdooring}

Formally speaking, the bug in the compiler does not take into account the \textbf{NOT} operator in context of equal checks. The legitimate use of \textbf{NOT} operator with equal (or difference) test is with Boolean values (meaning $a \in \mathbb{F}_{2}$ with $\mathbb{F}_{2} = \left\{0, 1\right\}$). This is not a big assumption since conditions, where negations are used, manipulate, most of the time, Boolean or equivalent values. With the compiler's bug, all the following codes are equivalent: $\forall a, b \in \mathbb{F}_2$, we have $a == b \iff !a == b \iff a ==\ !b \iff !a ==\ !b$ which, obviously, leads to issues. A more complex example can be illustrated with: $!\left(!a == b\right)$. In a non bogus environment, we have the equivalent conditions:
\begin{eqnarray*}
!\left(!a == b\right) & \iff & !\left(a \neq b\right) \\
\mbox{ } & \iff & a == b
\end{eqnarray*}

In the bogus environment of \textbf{ml} compiler, truth equation should change. It is due to the \textbf{NOT} operator which is not interpreted correctly. The first \textbf{NOT} is interpreted on the whole condition while it changes the jump instruction. From an equality, it is now an inequality. The second \textbf{NOT} (in the brackets) is not interpreted at all, which leads to the following conditions:
\begin{eqnarray*}
!\left(!a == b\right) & \iff & !\left(a == b\right) \\
\mbox{ } & \iff & a \neq b
\end{eqnarray*}

With this change in logic of program, we can replace a condition, normally not executed, into one which is now executed. Note the opposite operation is although correct: it is possible to change a condition which is executed to one which is no more executed. This is the main point to build the plot of a backdoor.

\subsection{Creation of the backdoor}

\begin{figure}[!h]%
    \centering
   \includegraphics[width=\columnwidth]{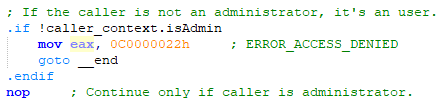}
   \caption{Simple backdoor using the compiler's bug.}
   \label{fig:exampleBDI}
\end{figure}

Using logic bug to change behavior of a targeted application is quite simple. We illustrate things with the use of access check to a resource reserved to administrator. For real instances, one can think to programs such as \textit{runas} or \textit{sudo} under UNIX subsystem. The goal is to introduce a flaw in a function supposed to check access rights of the calling process before continuing potential privileged functionalities. The simplest example could look like the one provided in figure~\ref{fig:exampleBDI}.\newline

With the previous code, if a regular user calls the function, it gains access to the privileged function. This is due to the fact that the \textbf{NOT} operator is not correctly interpreted by the compiler. Finally, \textbf{ml} compiles a condition close to "\textsf{.if caller\_{}context.isAdmin == 1}" which triggers the condition, so that the function is stopped. But the code in figure~\ref{fig:exampleBDI} is wrong for an operational backdoor. Indeed, if an administrator calls the function, because of the bug, the legitimate access is now refused. In this case, even the simplest checking test would found something goes wrong. If we introduce a backdoor, this one must keep the original behavior of the modified code. Inserting a new \textit{backdoor functionality} does not mean to remove already present functionalities.\newline

The solution comes with a small raise of complexity about the code to backdoor. Continuing the \textit{runas} example, this one can check, first, if the user belongs to a group, then if the couple user-name and password is correct. At that time, if the user belongs to the group of administrator, access can be guaranteed. A correct implementation would lead to check all these operations ones after the others. But, for optimizing things up, we propose to check is the provided user-name is administrator or just a regular user (implicitly supposing there is no other group except administrator or regular user). Finally, the check of provided password can be performed. The resulting code can be seen as the following.

\begin{figure}[!h]%
    \centering
   \includegraphics[width=\columnwidth]{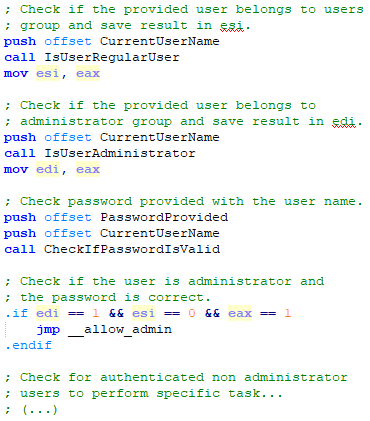}
   \caption{Check to let access to provided login/password.}
   \label{fig:OriginalCheckRunAS}
\end{figure}

The \textit{double} check of the user affiliation can be justified by the fact that non administrator user could perform some task in the coming code or something else, close to that explanation. The code in figure~\ref{fig:OriginalCheckRunAS} is the original code to backdoor. Of course, the main working space is in the condition. First, we are going to study the condition with a truth table, according to all possible outputs of values:\newline

\begin{tabular}{|c|c|c||c|}
\hline 
Administrator & User & Password correct & Result \\ 
\hline 
0 & 0 & 0 & 0 \\ 
\hline 
0 & 0 & 1 & 0 \\ 
\hline 
0 & 1 & 0 & 0 \\ 
\hline 
0 & 1 & 1 & 0 \\ 
\hline 
1 & 0 & 0 & 0 \\ 
\hline 
1 & 0 & 1 & 1 \\ 
\hline 
1 & 1 & 0 & ? \\ 
\hline 
1 & 1 & 1 & ? \\ 
\hline 
\end{tabular} 

\mbox{}\newline

Note the last two possibilities should never happen. Indeed, by definition, a user cannot be both administrator and regular user (ie: non administrator) and even if it was possible, there is no real definitive answers to provide here. In a way, these undefined states could be an interesting stop to exploit our compiler's bug... Technically, our goal is to get a regular user (ie: non administrator), with the correct password of the user account, being executed as an administrator, in addition to administrator users whose execution is still guaranteed. First, we are going to modify the condition in a manner it does not change the logic of the condition but it introduce our \textbf{NOT} operator. One can write:

\begin{figure}[!h]%
    \centering
   \includegraphics[width=\columnwidth]{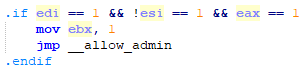}
   \caption{Change the condition (but not the logic) to introduce \textbf{NOT}.}
   \label{fig:NorBackdoorCondition}
\end{figure}

The code in figure~\ref{fig:NorBackdoorCondition} introduces the \textbf{NOT} operator but it does not change the logic of the condition from a formal point of view. Note that there is many ways, with the compiler's bug, to introduce this operator (for instance, \textsf{!esi == edi} since \textsc{edi} is supposed to be at one). This introduction is targeted on the \textsc{esi} register since this one is supposed to contain the Boolean value defining the belonging of the provided user to the group of regular users. This is in order to force the misinterpretation the real membership of the provided user at execution time.\newline

\begin{figure}[!h]%
    \centering
   \includegraphics[width=\columnwidth]{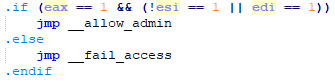}
   \caption{Trapped condition.}
   \label{fig:BackdoorCondition}
\end{figure}

The logic of the condition is to only accept what is defined as true to let privileged functionalities being executed. With just the last modification, no user (administrator or not) would be executed. Indeed, the condition is now only valid if the user is identified as being membership of both administrator and non administrator group. This is due to the use of \textbf{AND} operators in the condition. To remove this obstacle, we can define condition in other order and by the use of negative logic. Everything which is identified not to be an authenticated user will be rejected, allowing the execution to continue otherwise. Of course, because the undefined state is not supposed to happen in theory, we do not implement it in our code. A possible solution can be seen in code written in figure~\ref{fig:BackdoorCondition}.\newline

Condition in figure~\ref{fig:BackdoorCondition} is conform with the logic of the original condition in figure~\ref{fig:OriginalCheckRunAS}. It first checks whether the password is correct or not. If this one is not compliant, the code jumps directly to the failure part of the function. The justification of this first check lies in the need to avoid unnecessary checks if the password is invalid. Then, if the password is correct, the condition checks if the provided user is either an administrator or a regular user. This double check is to \textit{officially} avoid \textit{undefined} membership status of users...\newline

The reality of the compiler's bug forces the condition to be interpreted differently. Actually, the sub-condition \textsf{(!esi == 1)} is interpreted as \textsf{(esi == 1)}. Under this condition, it allows an administrator or a regular user, provided the password linked to its account, to get access to privileged functionalities. The following array resumes the truth table really compiled by the \textbf{ml}.\newline

\begin{tabular}{|c|c|c||c|}
\hline 
Password correct & User & Administrator & Result \\ 
\hline 
0 & 0 & 0 & 0 \\ 
\hline 
0 & 0 & 1 & 0 \\ 
\hline 
0 & 1 & 0 & 0 \\ 
\hline 
0 & 1 & 1 & 0 \\ 
\hline 
1 & 0 & 0 & 0 \\ 
\hline 
1 & 0 & 1 & 1 \\ 
\hline 
1 & 1 & 0 & 1 \\ 
\hline 
1 & 1 & 1 & 1 \\ 
\hline 
\end{tabular} 

\mbox{}\newline

The antepenultimate and the penultimate lines define the expected result. Either an administrator or a regular user --- correctly authenticated --- checks the condition. This is the design of the backdoor we introduce. It allows one more \textit{hidden} case to get access to privileged functionalities while keeping the original behavior from the condition. Remaining the case of the last line, this one is not really relevant, since it is theoretically not expected a single user can be a member of two groups at the same time. And, even if it could happen, officially, the condition would give access if there is admin rights. To some degree, condition made a choice to privilege admin users, whatever they belong to another group at the same time... A perfectly sustainable choice.\newline

At the difference of \cite{BackDoorBugCompiler}, our code does not appear as \textit{needlessly complicated}. It is perfectly justifiable without betray its unexpected design. One malicious author could insert such a code as a patch and justify it by optimization purposes, as we did in this paper. Another difference lies in the flexibility of our backdoor. It can be introduced in any condition, changing the logic flow of this one without removing the original behavior of the condition. To perform the modification, one can follow these steps:

\begin{enumerate}
	\item Define the original truth table of the condition~;
	\item Write the expected truth table for the new condition~;
	\item Build a legitimate condition which follows the last truth table by the use of \textbf{NOT} operator in an equal test condition to exploit the bug.\newline
\end{enumerate} 

\section{Correction of the bug}

Once the bug has been discovered, and because it is potentially critical, it has been transferred to Microsoft under CVE number CVE-2018-8232\footnote{\url{https://portal.msrc.microsoft.com/en-US/security-guidance/advisory/CVE-2018-8232}}. Since the bug is clearly identified, detection of specific patterns in conditions is enough to detect potential issues. From that point, many possibilities are doable. First, the compiler could display warning or error messages to prevent compilation of such code. If it does not fix the problem, it allows developers to be informed about it and to avoid such a dangerous use. The second possibility is to increase the volume of generated code for such configuration. As presented in Section~\ref{sec:definedproblem}, the main problem comes from the lack of a temporary variable. These two solutions are presented in the current section.\newline

% Once the bug has been discovered, and because it is potentially critical, it has been transferred to Microsoft. After several mail exchanges, Microsoft has considered that if the bug exists in its compiler, this one is not critical since resulting vulnerabilities do not belong in their software but in those which have been compiled with their own compiler. In that sense, they did not provide to that bug a CVE number. When asking them about the potential impact of such a bug in third party software and the necessity to broadcast information, if Microsoft recognizes that the main mitigation is code reviews, they do not plan to correct it with more disclosure than any regular bug. It is regrettable that the disclosure must be performed in such a way since CVE exists and are designed for that case... More information about this exchange can be found at [XXXX].\newline

% With the compiler's bug, we are dealing with incorrect code generated in the following situation.
In addition, it was required to inform people about the necessity to review all their source-codes which used that compiler. To correct the bug, many solutions exist to do it. On the first hand, it would be about adding a warning when the compiler encounters such a construction (detected as previously bugger). This warning informs a potential flow in the construction when the negation must be performed with a temporary variable. On the second hand, the compiler can insert, by itself, this temporary variable to correct the logic. With the compiler's bug, we are dealing with incorrect code generated in the following situation.\newline

\begin{figure}[!h]%
    \centering
   \includegraphics[width=\columnwidth]{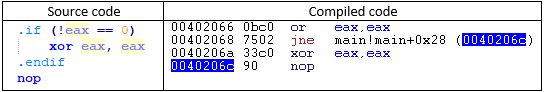}
   \caption{Initial code to correct.}
   \label{fig:CorrectionBase}
\end{figure}

Correction can be designed in the same way the C compiler did it in figure~\ref{fig:IdaDecompiled}. By saving original value of the value or by creating a new one, the \textbf{NOT} operation can be performed (explicitly by adding a test or just by using dedicated instructions). In that way, the compiled code could be generically corrected to something close to the code displayed in figure~\ref{fig:CorrectionBase}.\newline

\begin{figure}[!h]%
    \centering
   \includegraphics[width=\columnwidth]{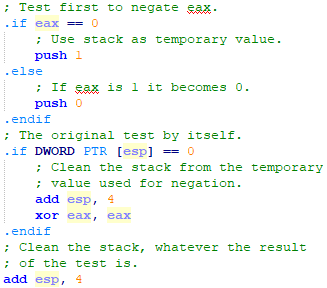}
   \caption{Proposition of solution to correct the bug.}
   \label{fig:CorrectionBase}
\end{figure}

The main idea is to save the negated value in the stack to use it directly in the original condition. Using stack is not a big issue since this is the regular procedure to create local values for functions. Note this temporary value used must be removed from the stack whatever the result of the condition was. It is in order to keep coherency for previous values already present in the stack.\newline

\begin{figure}[!h]%
    \centering
   \includegraphics[width=\columnwidth]{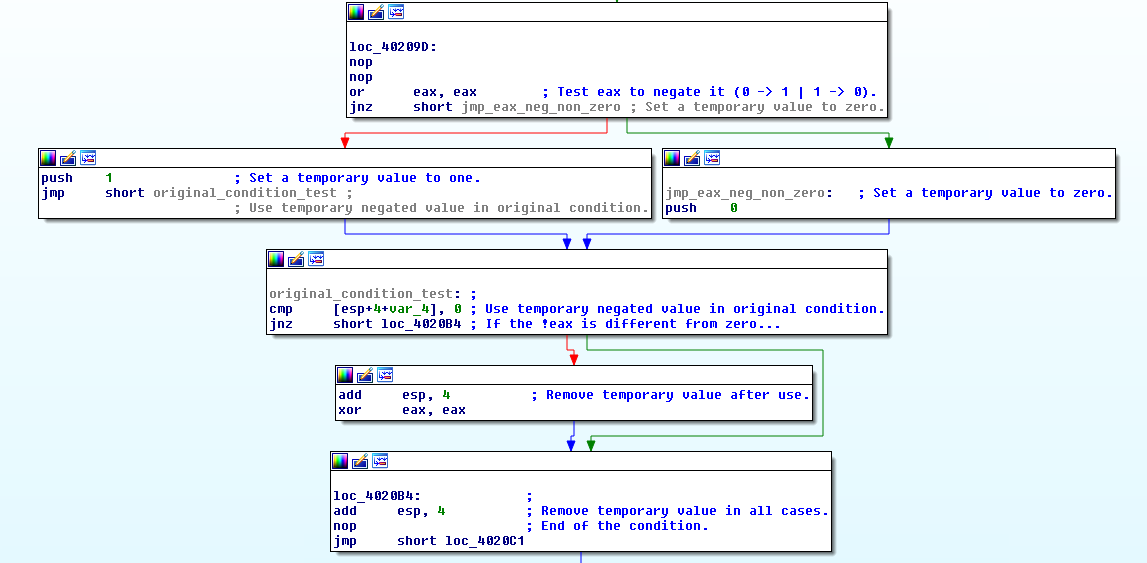}
   \caption{Decompiled correction for the proposed solution.}
   \label{fig:CorrectionBaseId}
\end{figure}

The decompiled version of the previous code follows requirement of the condition originally defined by the developer. The complexity cost is relatively small since operation on stack are common. No double code displayed in figures~\ref{fig:CorrectionBase} et \ref{fig:CorrectionBaseId} could be improved to something more efficient.\newline

More than correcting the bug in the compiler by Microsoft, a large audit of applications developed with that compiler is mandatory. Indeed, a compiler is a software responsible to generate other software. It means that other source codes have been compiled with it and these ones should be audited in order to find bogus constructions in conditions, on the first hand, and to confirm the original behavior of the software once this one has been recompiled (since condition flow could change) on the other hand. Taking into account that assembly language is used by some firmware builder, drivers developers or just for tenuous part of critical code, it can be complicated to update these software since some code can potentially not be updatable (lake of internet connection, code written in ROM for embedded devices, small portion of code used in a math library which is part of a bigger project...). Including the fact that ml is an old compiler, it represents a large number of software which have been made with it by a lot of people. Even if a CVE number has been assigned and a security update pushed, it could be hard to fix everything, especially if the source code of one software has been developed at former time by a company which has collapsed since that time. In such a case, a reverse engineering process should be made on compiled code to check that logic flow for each condition is designed to perform the task for those it has been designed.

\section{Conclusion}

This bug has been found during the development of specific processes dedicated to security. If there have always been debates over the usefulness and performance of assembly language relative to high-level languages, this one is still used in industry. Far to be the most common, the TIOBE index of programming language popularity ranks assembly language at 15 in February 2018\footnote{\url{https://www.tiobe.com/tiobe-index/}} and its use is growing these last years\footnote{\url{https://www.tiobe.com/tiobe-index/assembly-language/}}. Just from open-source world, one can find easily more than one thousandth projects using MASM in their source code\footnote{\url{https://searchcode.com/?q=masm&lan=34}}. We find assembly developers in the fields of compiler, firmware, security, real-time simulations or calculation and anti-malware engineering. All are critical fields and where programs run with high privileges and full access to user's data.\newline

After numerous exchanges with Microsoft team, a solution has been implemented to fix the bug. According to the previous section, several possibilities could be performed, including rejecting the compilation with an error or building the binary as C compiler does. But, because assembly language must give to the developer a full control of what it is compiled and in order to not break backward compatibility with existing software, Microsoft decided to correct the bug with the error message. Now, A2154 error with the message "syntax error in control-flow directive" is displayed when such construction in condition is used. From an operational point of view, recompiling existing codes with the new version of Visual Studio 2017 (July 2018) is enough to detect a potential use of the vulnerability.\newline

More generally, using a bug in a compiler to introduce a trap in a software is a nice and smooth manner to get access to a targeted system in the stealthiest way. From a specific compiled version dedicated to a specific target to a wildly used open-source software, possibilities are endless. Undetectable for humans, perfectly justifiable for malicious developers, thin and efficient, this type of backdoor is designed for long term and high efficiency. Detectable by automatic tests process, these one must be calibrated to record and take care of suspicious signal or undefined states. Of course, example provided in this paper is for illustration purpose and it could be possible to detect it with dedicated tests. But, with no prior knowledge a backdoor has been inserted, detection by specific test procedure can be almost impossible to perform. Note that, even if tests find something tendentious, auditing source code would not be enough to fix the problem and it would be even worth to identify the \textit{bug} as a malicious trap inserted on purpose... Reverse engineering is the only solution to understand and fix bug (both from the targeted software where the trap has been set and the bugged compiler), but it is not a mass sport and not a basic skill earned by each developer. Note that it requires that developers suspect bugs come from compiler and not their own code to first be able to fix things. A state of mind which is far to be common if you are not aware it could be possible.\newline

From an operation point of view, such a backdoor is almost perfect. Especially with the bug provided here, attacker has the ability to insert new possibilities in any critical condition targeted without modifying original behavior of the condition. Bonus, attacker can stop diffusion of the attack by correcting bug in the compiler, removing the trap for new compiled version without changing targeted software's source code. Another bonus lies in the difficulty to update critical pieces of code, most of the time in firmware and other close to hardware or kernel components. Last attacks such as Spectre \cite{Kocher2018spectre} and Meltdown \cite{Lipp2018meltdown} demonstrate how hard it could be to change firmware in critical pieces of software. Not using such extreme examples, thinking about pieces of software in firmware, written in assembly, sometime  years ago, by companies which could not exist any more, could make impact of such backdoor very critical.\newline

Of course, the goal of this paper is to report this vulnerabilities to make it patched as soon as possible. But it aims to explain how to find such a bug, make it visible to any developer and alert about how important damage could be if it was exploited by malicious developers. More important, it underlines the necessity to not believe blindly that open-source projects are secure. Even if they are correctly audited by experts. In the same philosophy than \cite{BackDoorBugCompiler}, our work is different since it does not require to use obfuscation to hide the trap, which make our perfectly justifiable. Compiler is a milestone in the building procedure of an application and it must be not underestimated as a potential source of vulnerability. This is not new since Ken Thompson \cite{Thompson} did it in 1984, by modifying the compiler and then compiling a trapped code. Here, we exploit a zero-day vulnerability in the compiler to do the job.\newline

% This has two advantages. First, obviously, no amount of inspection ? nor even full formal verification ? of the source code will find the problem. Second, the bug can be targeted fairly specifically if our target audience is known to use a particular compiler version, compiler backend, or compiler flags. It is impossible, even in theory, for someone who doesn't have the target compiler to discover our backdoor.

%Honestly, there is not so many widely used projects only written with MASM assembly. Most of the time, it's for critical application in embedded system or for compiler projects. But it's still possible to design things such as this compiler issue can be exploitable.\newline

%Injection of a backdoor in a regular process can be performed if and only if a malicious developer has access to modify source of the software. 

%Written in the middle of a given source code, this type of bug can provide to the caller of a dedicated function strategic advantages while being able to not be detected by reviewer of code. One can imagine a simple code supposed to check access to a specific resource and giving access to this one only if the user is correctly authenticated. In such a case, reading the code would not throw any objection, but, at execution time, access would be guaranteed to unexpected users. 

% Différence avec les autres : on n'a pas besoin de code bizarre, au contraire.
% Parler d'une solution de correction.
% Parler d'une méthode plus générique array[++i] = ++i.

\bibliographystyle{unsrt}
\bibliography{example}

%\section{Annexes}

\end{document}